\documentclass{article}
\usepackage[a4paper, total={6in, 8in}]{geometry}
\usepackage{amsmath}
\usepackage{amssymb}
\usepackage{nccmath}
\usepackage{algorithm}
\usepackage{algpseudocode}
\usepackage[
backend=biber,
style=numeric,
sorting=none,
maxbibnames=4
]{biblatex}

\renewbibmacro{in:}{}

\DeclareSourcemap{
  \maps[datatype=bibtex]{
    \map{
      \step[fieldset=abstract,   null]
      \step[fieldset=language,   null]
      \step[fieldset=urldate,    null]
      \step[fieldset=issn,       null]
      \step[fieldset=copyright,  null]
      \step[fieldset=shorttitle, null]
      \step[fieldset=keywords,   null]
      \step[fieldset=note,       null]
      \step[fieldset=month,      null]
      \step[fieldset=url,        null]
    }
  }
}

\usepackage{bbm}
\usepackage{svg}
\usepackage{caption}
\usepackage{graphicx}
\usepackage{xcolor}
\usepackage[hidelinks]{hyperref}
\usepackage{float}
\usepackage{authblk}

\title{Identifying structural design principles shaping the computational abilities of recurrent neural networks}
\date{\today}

\author{Tom Talpir}
\author{Elad Schneidman}
\affil{Department of Brain Sciences, \\ Weizmann Institute of Science, Rehovot, Israel}

\begin{document}

\maketitle
\begin{abstract}
Understanding how the architecture of neural networks shapes the computations they carry is a central challenge in neuroscience and machine learning. While specific circuit architectures have been linked to particular network computations and theoretical bounds on expressivity of broad classes of networks have been found, we are still missing general principles connecting the structure of finite networks to their computational capabilities. Here, we characterize the computational abilities of recurrent neural networks as a function of their connectivity by training a large collection of different networks to compute a large set of Boolean functions. For small networks, we constructed the complete ``catalogs'' of network-function performance, which revealed that computational capacity varies widely across architectures and that most networks show poor performance, and most functions are hard to compute. However, we show that having local 2- and 3-cycles in a network strongly enhances its computational ability, and networks with such cycles are often the minimal architectures that can solve particular functions. We further show that a small set of structural statistics accurately predict networks' performance. Extending our analysis to large networks showed that typical networks fail even to approximate a randomly selected function. Surprisingly, adding a small number of sparsely connected biologically-inspired interneurons to the network dramatically increases computational capacity. As in small networks, adding short cycles improved networks' capacity, outperforming acyclic or reachability-matched controls. Thus, our results identify local cycles as design principles linking neural connectivity to computational power, and offer a general framework to explore structure-function relations in computing networks.  

\end{abstract}

\newpage 

\section{Introduction}
The architectures of real-world networks often exhibit structures that reflect their functional design or their development. For example, power-law or log-normal degree distributions are commonly found in biological, social, or linguistic networks, conferring robustness to errors and failures \cite{barabasi_power_law, barabasi_error_tolerance}. These structures can be explained by different, relatively simple, developmental processes, such as preferential attachment, spatial branching and competition \cite{kaiser_spatial_2004, kaiser_simple_2009}. Similarly, small sub-network ``motifs'' are much more frequent than expected by chance, in  biological and engineered networks \cite{alon_milo_motifs}, and have been associated with specific computational roles \cite{alon_FF_motif}. The remarkable computations carried by biological neural networks \cite{hafting_microstructure_2005, mante_context-dependent_2013} and artificial ones \cite{lecun_deep_2015, jumper_highly_2021}, and their ability to learn in supervised \cite{Krizhevsky_imagenet}, unsupervised \cite{goodfellow_GAN}, or reinforcement learning \cite{silver_general_2018, mnih_human-level_2015} makes them particularly appealing systems to study the relations between the structure of networks and their computational ability. 

The reconstructions of detailed connectivity maps of neural circuits, or ``connectomes'' \cite{dorkenwald_neuronal_2024, cook_whole-animal_2019, white_structure_1986} now enables the exploration of the relationship between the fine structure of neural circuits and the function that they may perform \cite{beiran_prediction_2025, litwin-kumar_constraining_2019}. While in a few cases the relations between architecture and function have been mapped and understood \cite{kim_ring_2017, lyu_building_2022, briggman_wiring_2011}, most circuits do not have clear structural design or symmetries that reflect function, and while simulating particular connectomes may reflect on circuits' function in specialized cases \cite{wanner_whitening_2020, lappalainen_connectome-constrained_2024}, these do not offer a general understanding of these mappings. Surprisingly, generative models of neural circuit architectures suggest that they may be explained by a small number of biological and physical features \cite{betzel_generative_2017, haber_schneidman_architectural_features, haber_schneidman_connectome, richter_building_2025}, but it is unclear how these building blocks affect which computational tasks a network can learn to perform, or how certain structural properties enhance computational performance.

Compared to the typically recurrent and sparse architectures of biological neural networks, and their energetic efficiency, artificial neural networks, that were inspired by real neural networks, are typically very different. In particular, many of these networks rely on highly feed-forward architectures, and their key structural properties are the depth of the network and width of layers, the activation function of the neurons, and specialized layer structures, such as fully connected layers, convolution ones, or attention layers. Foundational theoretical work has shown that Feed-forward neural networks with exponential width can approximate any continuous function \cite{hornik_universal_approx}, as well as lower and upper bounds on what a circuit can compute \cite{hastad_switching_lemma}, which have even been extended to some models of neural networks \cite{parberry_circuit_complexity}. Moreover, depth has been shown both empirically and theoretically to play a role in determining the power of neural networks \cite{shamir_eldan, bengio_depth_effect, Krizhevsky_imagenet}. Theoretical frameworks such as PAC learning \cite{valiant_PAC_learning} attempt to answer what artificial neural networks can learn to approximate. However, these results, which are mostly asymptotic in nature, do not immediately translate into how local properties of finite networks determine their power and expressivity. This is particularly true for recurrent networks, which are prevalent in many neural circuits \cite{luo_RNN_in_nervous_system, douglas_martin_neocortex_recurrent_circuits}, and while their computational power has been explored both theoretically and empirically \cite{maass_real-time_2002, sussillo_generating_2009, sussillo_opening_2013}, these approaches characterize computation in terms of dynamics, not connectivity structure. 

For a particular class of neurons, namely combinatorial inhibitory threshold-linear networks, the nature of network dynamics can be proven to be predicted from the network’s structure  \cite{curto_graph_rules, curto_graphical_2025}. However, these results only focus on the nature of network dynamics, without addressing the challenge of learning detailed computational tasks, and are yet to be generalized to other classes of neurons and networks. Analysis of spiking neural network models have identified simple architectural features that predict the functional similarity of network architectures \cite{haber_schneidman_architectural_features}. Ideally, we would like to combine such approaches to form a general prediction of function from structure. 

While extensive work in graph theory and network science has characterized how the structure of a network shapes information flow, robustness, and the dynamics that unfold on networks \cite{albert_statistical_2002, newman_structure_function_nets}, the question of which computations a specific graph topology can actually carry has received less attention. We take a directed approach and investigate how the connectivity of recurrent neural networks shapes the computational tasks they can perform --- by training a large ensemble of network architectures on a large and rich set of computational tasks and evaluating their performance (Fig.~\ref{fig:Figure1}). The space of possible tasks is both vast and hard to define precisely \cite{sipser_TOC}, and so we focus on Boolean functions as a comprehensive task space: they are diverse enough to capture a wide range of computational demands, and the ability to learn arbitrary Boolean functions provides a stringent test of network capacity, including the ability to fit effectively random structure. Because the number of different connectivity graphs for $N$ neurons and the number of Boolean functions on $N$ bits grows exponentially with $N$, exhaustive analysis is feasible only for small networks, where we enumerate all architectures and tasks. For larger networks, we resort to sampling over both network architectures and functions. Combining the exhaustive analysis of small networks with sampled exploration at larger scales, we seek general principles linking network connectivity to learning ability, and to understand how structural constraints on dynamics and information flow shape the structure–function relationship in recurrent networks.

\section{Results}
\begin{figure}[H]
    \centering
    \includegraphics[width=\linewidth]{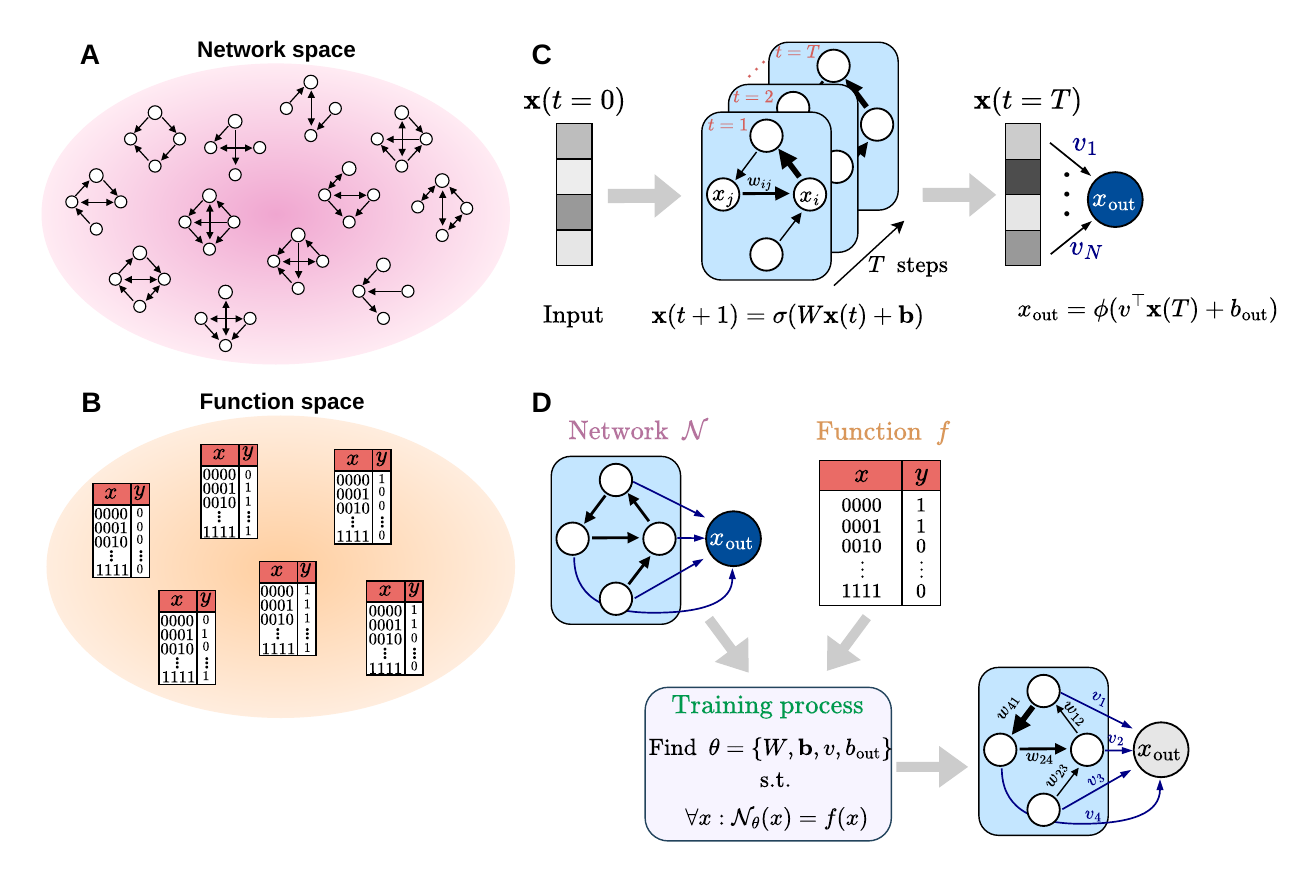}
    \caption{\textbf{Mapping the space of network architectures that can learn to compute individual Boolean functions.} \textbf{(A)} Illustration of the space of different recurrent neural network connectivity maps, which consists of all $2^{N(N-1)}$ directed graphs on $N$ nodes; Shown here are samples of the case of $N=4$. \textbf{(B)} Illustration of the space of Boolean functions $f:\{0,1\}^N \rightarrow \{0, 1\}$, each represented as a binary truth table of length $2^N$, which implies there are $2^{2^N}$ such Boolean functions; Again, samples here show the case of $N=4$. \textbf{(C)} Schematic of the dynamics and computation carried by our recurrent neural networks for the case of $N=4$ neurons: A binary input pattern initializes the network state $\mathbf{x}(t=0)$, which evolves for $T$ steps via the recurrent update rule (Eq.~\ref{eqn:Eq1}). The final state $\mathbf{x}(T)$ is read out by a single output neuron (Eq.~\ref{eqn:Eq2}). \textbf{(D)} Schematic of the training of networks: for each pair of network architecture $\mathcal{N}$ and target Boolean function $f$ (represented by its truth table), we optimize the parameters $\theta = \{W, \mathbf{b}, v, b_\mathrm{out}\}$ such that the network correctly computes $f$ on every Boolean input $x$, i.e., $\mathcal{N}_\theta(x) = f(x)$ using BPTT (see \nameref{Methods}).}
    \label{fig:Figure1}    
\end{figure}

We consider all the different connectivity maps of networks of $N$ neurons. Each such wiring map, or topology, is a directed graph specifying which neurons connect to which, with self-connections excluded; there are $2^{N(N-1)}$ such graphs on $N$ neurons. The dynamics of a neural network with a particular topology (Fig. \ref{fig:Figure1}C) is defined by the edge weights ``on" this topology, namely the $N \times N$ real matrix $W$, where $W_{ij}$ is the strength of the connection from neuron $i$ to neuron $j$, and $W_{ij}=0$ wherever no edge is present.

The input to a network sets the initial state of each of the neurons at time $t=0$, namely, $x_j(t=0)$, and at each time step $t$, each neuron updates its state according to
\begin{align}
    \label{eqn:Eq1}
    x_j(t+1) = \sigma ( \sum_{i \neq j } W_{ij} \cdot x_i(t) + b_j  )
\end{align}
where $b_j \in \mathbb{R}$ is a bias term, and $\sigma$ is a nonlinear activation function. After $T$ steps, the state of the network, $\mathbf{x}(T)$ is read by a readout neuron, and so the output of the network is given by  
\begin{align}
    \label{eqn:Eq2}
    x_{\text{out}} = \phi(\sum_{i=1}^N v_i x_i (T)  + b_\text{out})
\end{align}
where $v_i$ is the readout weight between neuron $i$ and the output neuron, $b_\text{out} \in \mathbb{R}$ is a bias term, and $\phi(\cdot)$ is the Heaviside threshold function where $\phi(z) = 1$ if $z \ge 0$ and $0$ otherwise.
    
We train each of the different networks to compute each of a large set of different target functions. Since the space of functions we could consider is essentially unbounded, we seek a diverse set of functions that is rich enough to reflect a network's computational capacity. We therefore use the full set of Boolean functions $f:\{0,1\}^N \rightarrow \{0, 1\}$, so each of the $2^N$ possible binary inputs of length $N$ is mapped to a binary output. There are $2^{2^N}$ such Boolean functions on $N$ bits, which provide a large set of computational tasks of different complexity and dependencies. Moreover, since Boolean functions are well-studied and have a rich mathematical foundation \cite{odonnelBoolFunctions}, we can leverage analytical tools to quantify function complexity and devise efficient sampling strategies (see \nameref{Methods}).

We consider different values of $N$ and explore different classes of the neuronal activation functions, which may impact the computation that a network can perform or the convergence of the network across various tasks \cite{activation_functions_survey}. We used both Sigmoid and ReLU activation functions and found no significant effect on learning performance, and therefore focus on the Sigmoid function (see Supplementary Figure~\ref{fig:suppfig_relu_vs_sigmoid}). The running time of the network $T$ is another important factor shaping a network's computational ability; here we focus on $T=3$ and include a preliminary analysis of other runtimes in Supp. Fig.~\ref{fig:runtime}.

As the number of possible network topologies and the number of Boolean functions prohibit simulating all networks for large $N$, we begin by exploring small networks, where we can exhaustively study all possible network architectures and evaluate their computational abilities, and use the insights from these networks to later study larger networks, where we have to use sampling techniques.

\subsection{Computational capacity varies dramatically across networks}
We evaluate the performance of all $2^{N(N-1)}$ possible networks for $N=3$ and $N=4$, where each network is trained separately to compute each of the $2^{2^N}$ Boolean functions over 3 or 4 bit inputs. Every network-function pair, $(\mathcal{N}_i, f_j)$, is trained for 500 epochs using Backpropagation-through-time (BPTT) \cite{BPTT} (Fig.~\ref{fig:Figure1}D, see also \nameref{Methods}). Since our goal is to assess whether a network can express a function, rather than its ability to generalize from limited data, the network is trained on the full truth table containing all $2^N$ input-output pairs, and no train/test split is used. To account for variability due to random weight initialization and reduce the chance of falsely classifying a network as unable to learn a function due to optimization failure, each network–function pair is trained independently across 10 random initializations.

\begin{figure}[p]
    \centering
    \vspace{-1.5em}
\includegraphics[width=\linewidth,height=0.85\textheight,keepaspectratio]{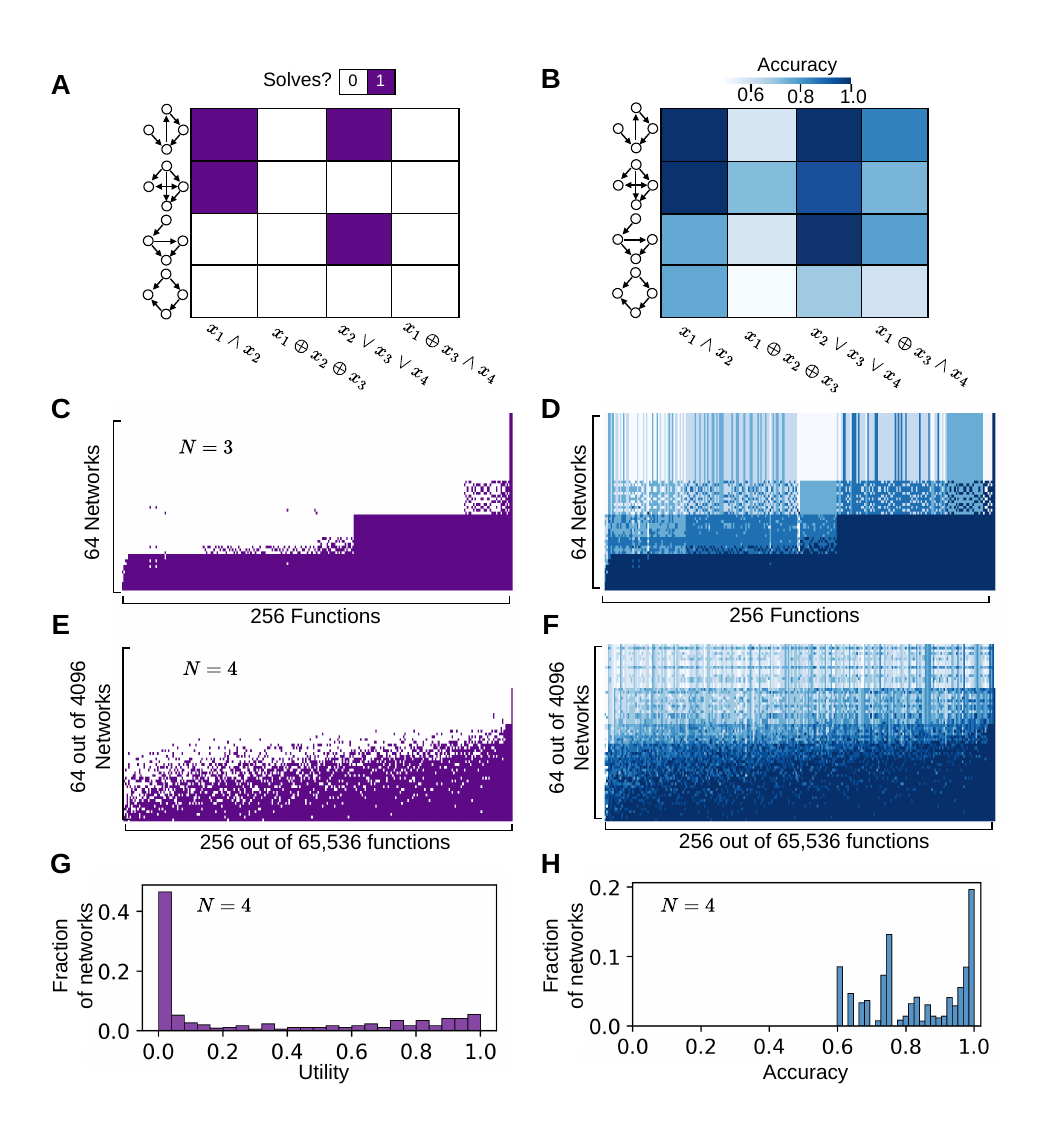}
    \vspace{-2.5em}
    \caption{\textbf{Catalog and Approximation matrices show that networks' computational capacity is widely distributed, but most show poor performance.} \textbf{(A)} A small part of the Catalog Matrix $\mathcal{C}$ for networks of size $N=4$; Here we show the values of $\mathcal{C}_{ij}$ for 4 example networks and 4 functions, which equals $1$ if network $i$ successfully learned to compute function $j$ and 0 otherwise. \textbf{(B)} A small part of the Approximation Matrix $\mathcal{A}$ for $N=4$; $\mathcal{A}_{ij}$ equals the highest accuracy network $i$ achieves on function $j$. \textbf{(C)} The full Catalog matrix for $N=3$; rows are sorted by how well networks performed (i.e., Utility), and columns by how difficult the functions were to learn (i.e., how many networks solved them). \textbf{(D)} The full Approximation matrix for $N=3$; sorting as in (C). \textbf{(E)} A subset of the full $N=4$ matrix entries, showing values for randomly selected 64 networks and 256 functions; sorting as in (C). \textbf{(F)} Same as (E), but for the Approximation matrix. \textbf{(G, H)} The distribution of Utility and Accuracy (respectively) over the full set of $N=4$ networks and Boolean functions.}
    \label{fig:Figure2}    
\end{figure}

For each such pair, we record whether network $\mathcal{N}_i$ successfully learned to compute function $f_j$ in a ``Catalog Matrix'' $\mathcal{C}$ , where $\mathcal{C}_{ij}=1$ indicates that network $\mathcal{N}_i$ learned to compute the correct output for all $2^N$ inputs of $f_j$ in at least one trial, and 0 otherwise (Fig.~\ref{fig:Figure2}A). We then quantify the computational ability of each network by the fraction of Boolean functions it learns to compute perfectly,
\begin{align}\label{eq:Utility_Eq}
    \text{Utility}(\mathcal{N}_i) = \frac{1}{2^{2^N}} \sum_{f_j \in \mathcal{F}} \mathcal{C}_{ij}
\end{align}
where $\mathcal{F}$ denotes the space of all Boolean functions. For $N=3$, we can plot the full Catalog Matrix, comprising 64 networks and 256 functions (Fig.~\ref{fig:Figure2}C); Sorting rows by Utility and columns by the number of networks that solve each function reveals a clear division: a small cluster of high-performing networks that solve many functions, and a large majority that solve almost none.

Since perfect computation is a strict criterion, we also consider how well a network approximates a function, namely the fraction of inputs for which it managed to learn to produce the correct answer. Thus, the Approximation Matrix $\mathcal{A}$ gives another facet of the networks' ability, where each entry $\mathcal{A}_{ij}$ is the highest accuracy that $\mathcal{N}_i$ managed to learn on $f_j$ across the 10 trials (Fig.~\ref{fig:Figure2}B). We then score each network by its average accuracy across all Boolean functions:
\begin{align}
    \text{Accuracy}(\mathcal{N}_i) = \frac{1}{2^{2^N}} \sum_{f_j \in \mathcal{F}} \mathcal{A}_{ij}
\end{align}
Sorting the $N=3$ Approximation Matrix by the same ordering as the Catalog Matrix (Fig.~\ref{fig:Figure2}D), reveals that while many networks fail to perfectly compute many functions, they nonetheless are capable of approximating many of them:  many networks reach high accuracy across much of the function space, reflecting that there is a substantial gap between approximation and exact computation.

We next computed the full Catalog and Approximation matrices for $N=4$; due to their size (4096 networks over 65536 functions), we present only a randomly sampled subset (Fig.~\ref{fig:Figure2}E-F). Sorting networks by their Utility again reveals heterogeneity in computational performance, with some networks failing to compute virtually any of the functions, whereas others learn to compute nearly all functions. With the exception of the two constant functions ($f(x)=0$ and $f(x)=1$), which all networks learn to compute, most functions are solved by only a small fraction of networks. Compared to $N=3$, the $N=4$ matrices show more intricate  structure, and sorting networks by Utility no longer yields a clean separation into distinct performance classes. We further find that most networks perfectly compute only a small number of functions (Fig.~\ref{fig:Figure2}G), while most networks achieve high accuracy on the majority of functions (Fig.~\ref{fig:Figure2}H). We also asked whether embedding networks into a low-dimensional space, where proximity of networks would correspond to similarity in the functions they learned to compute, would reveal interesting functional relations between them. While different embeddings could give different answers, Multi Dimensional Scaling \cite{cox_multidimensional_2008} and PCA-based embeddings did not expose meaningful structural features (Supp. Fig.~\ref{fig:SuppFig_CatMat_Distances}). 

Thus, while many of the networks turn out to be of limited capacity, and many functions are hard to compute, some networks still show high capacity, and so we next ask which features of a network's topology shape its computational ability. 
\subsection{Identifying specific connectivity patterns that shape the computational abilities of networks}
We proceed to examine how networks' connectivity shapes computational abilities, noting that if two networks $\mathcal{N}_{k}$ and $\mathcal{N}_{l}$ are identical up to a permutation of neuron labels then they have identical computational abilities, since a solution for one maps directly to the other by relabeling. We therefore group these \textit{isomorphic}  networks into equivalence \textit{network classes}. Similarly, Boolean functions can be grouped into equivalence classes based on permutations of input bits, which we call \textit{function classes}.

\begin{figure}[H]
    \centering
    \vspace{-4.5em}
    \includegraphics[width=\linewidth]{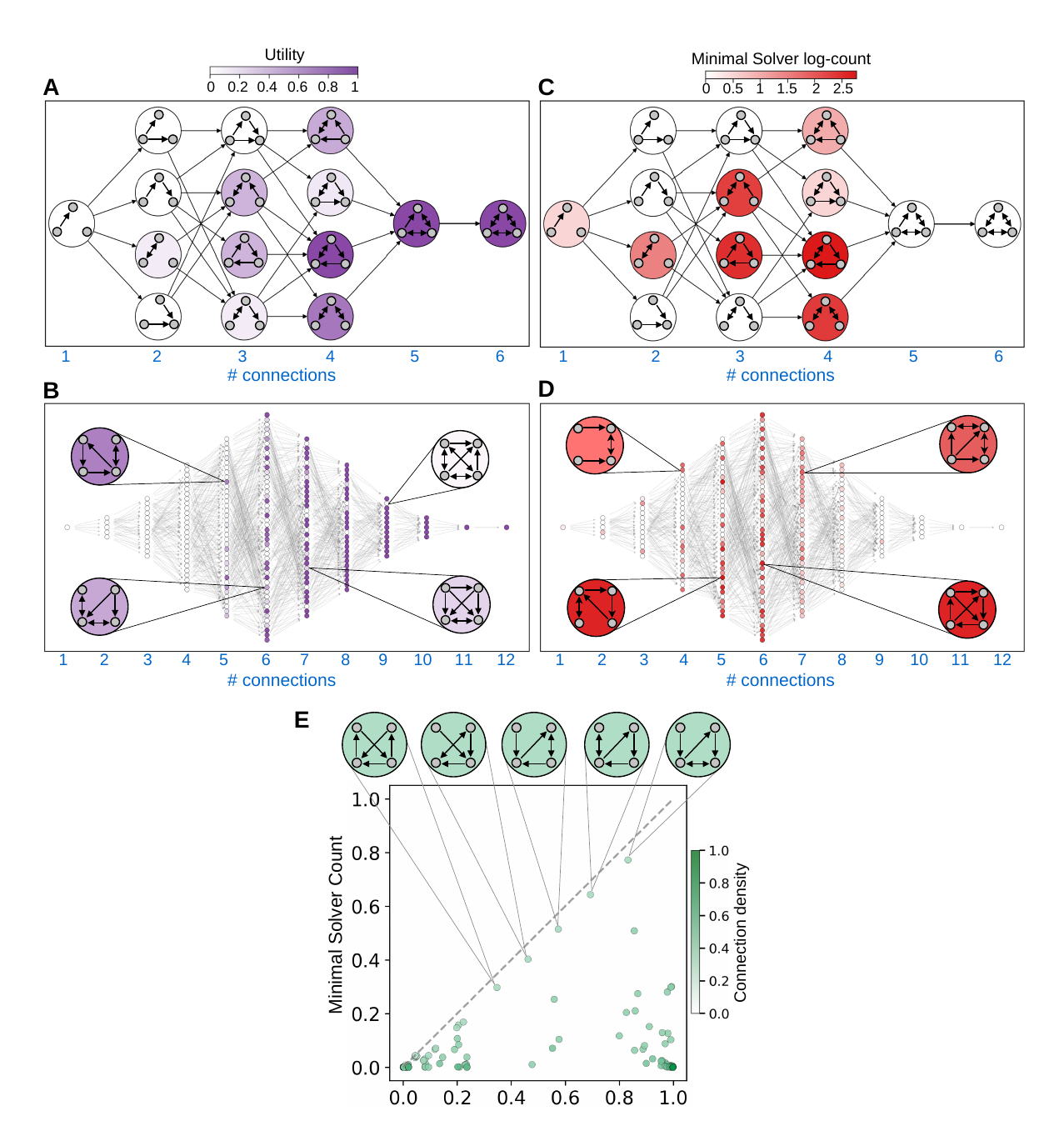}
    \vspace{-3em}
    \caption{\textbf{Hierarchical organization of networks by their connection maps reveals non-monotonic computational abilities and local connectivity structures that shape it}. (\textbf{A}) A tree-like organization of network architectures for $N=3$, where each node represents a network class that consists of all networks that are equivalent up to node label permutation. Going from left to right, nodes are connected as if adding a single connection to a network in the left class transforms it to the class on the right. Network classes are thus ordered from left to right by the number of connections in them, and are colored by the Average Utility of all networks in that class. (We omitted the empty network which can only solve the two constant functions). (\textbf{B}) Same as (A), but for $N=4$. (\textbf{C-D}) Network classes are ordered as in (A-B), but here the color of each node reflects the log of the number of function classes for which that network class is a minimal solver. Specific examples of classes are shown as ``zoom in'' to reflect particular class architectures that show particularly high or low performance compared to classes with similar connectivity. (\textbf{E}) For each one of the classes from (B) and (D) we plot the Utility of the class vs. its Minimal solver value. Dashed line shows the achievable bound, and the special classes that show high Utility and nearly maximal Minimal solver values are "zoomed in". }
    \label{fig:Figure3}    
\end{figure}

We organize network classes into a tree-like structure, where each node represents a network class, and the edges connect network classes that are the result of adding (or removing) a single connection to a network from that class (or removing one). The root of the tree is the class of networks with a single connection, and moving from left to right along the tree corresponds to progressively denser networks, with each level adding one connection. We plot these trees for $N=3$ and for $N=4$, coloring each node by the Utility of its network class --- the fraction of function classes that networks in that class can perfectly compute (Fig.~\ref{fig:Figure3}A-B). Notably, connection count alone does not determine performance: some sparser network classes achieve higher Utility than denser ones at the same level of the hierarchy. A preliminary inspection suggests that network classes containing short recurrent cycles seem to achieve high Utility.

We further quantify the ability of each class by asking for which functions a given network is the minimal one of all the networks that can compute that function. To capture the idea of structural necessity, we introduce the notion of \textit{Minimal Solvers}: A network class is a Minimal Solver for a specific function class if no ancestor of that node in the hierarchy tree (i.e., no sparser network obtained by pruning connections) can also solve that function. In other words, a Minimal Solver network has no redundant connections with respect to that function and its topology is the simplest possible that supports the computation. Then, for each network class we count the number of function classes for which it is a Minimal Solver, and visualize this on a similar hierarchical tree (Fig.~\ref{fig:Figure3}C-D). This analysis again highlights the benefit of short recurrent cycles: as networks with 2- or 3-cycles are Minimal Solvers for a broad range of functions, suggesting that recurrent motifs are not merely helpful but structurally essential for many computations.

Comparing the Utility score and Minimal Solver score across all $N=4$ network classes reveals several distinct behaviors (Fig.~\ref{fig:Figure3}E): The many networks with low Utility are typically not Minimal Solvers for any function, whereas some high-Utility networks also have a Minimal Solver score near zero, implying that sparser networks can replicate their solutions, and their additional connections are redundant. A small subset of network classes stand out by achieving Minimal Solver score that is nearly as high as their Utility score (which is an upper bound on their minimal solver score). Thus, these network classes are structurally efficient, as they solve many functions, and for nearly all of them, no simpler network suffices. Notably, these are precisely the network classes whose topology contains short recurrent cycles, which stood out for each of the performance metrics above. Taken together, we suggest that local recurrent connectivity not only broadens a network's functional repertoire but does so in a structurally irreducible way. We therefore turn to ask how well we can predict the computational ability of a network directly from its structure.

\subsection{Predicting computational abilities of networks from their structure}
\begin{figure}[H]
    \centering
    \vspace{-4em}
    \includegraphics[width=\linewidth]{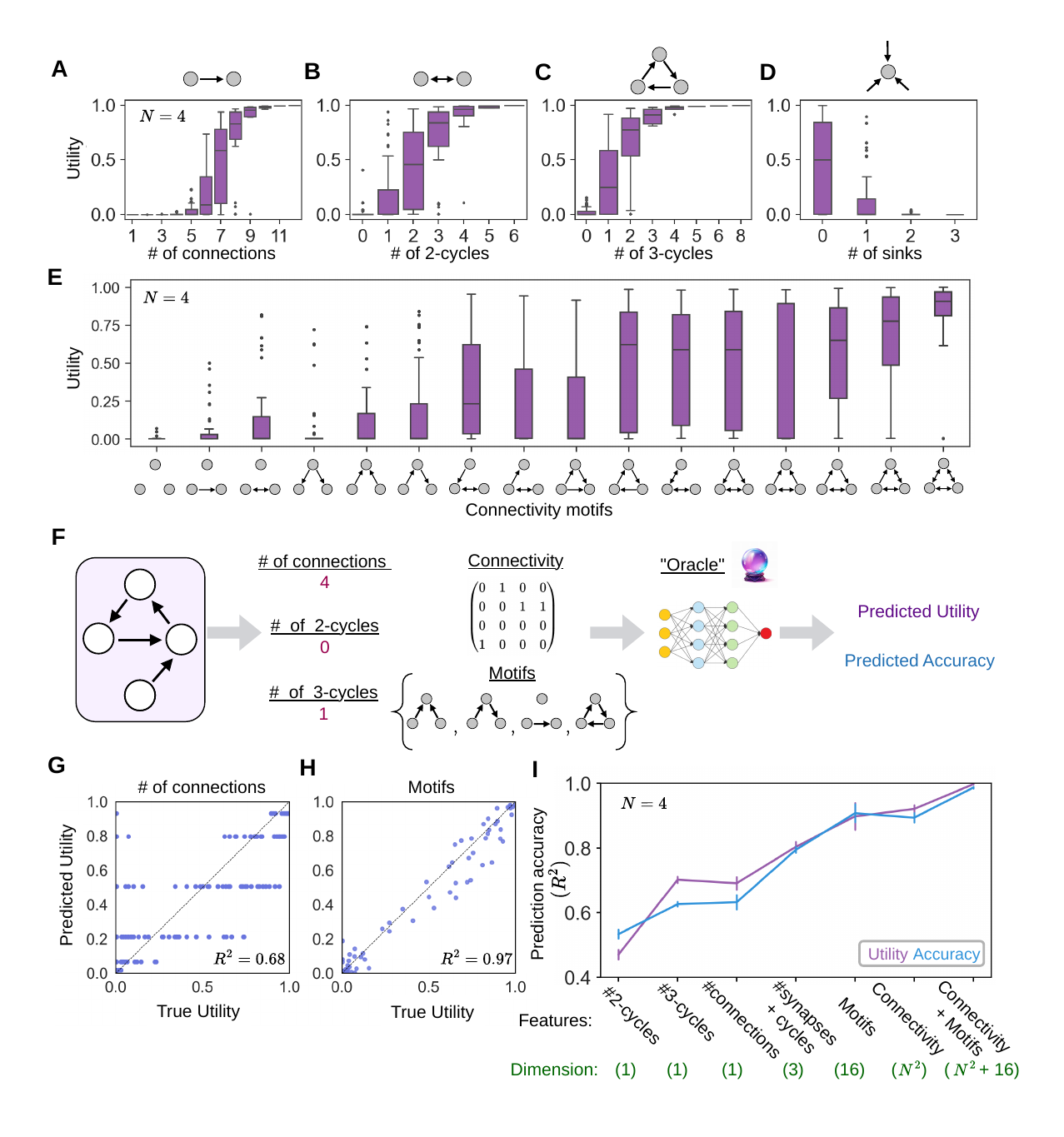}
    \caption{\textbf{Predicting network performance from structural properties of $N=4$ networks}. \textbf{(A-D)} The Utility of networks is shown as a function of their number of connections, number of 2-cycle, number of 3-cycle, and number of sinks. Median Utility per x-axis value is shown by a black horizontal line, box designates 25-75 percentile values, whiskers extend to the 5-95 percentiles, and outliers are shown as individual dots. \textbf{(E)} Utility values of networks containing each of the 16 non-isomorphic 3-node directed motifs. \textbf{(F)} An illustration of the prediction of a network's capacity based on its structural features, by a multilayer perceptron (``oracle'') whose weights were optimized to predict Utility or Accuracy. \textbf{(G-H)} Predicted Utility is shown vs. the measured one, for $N=4$ networks, where the features that were used by the oracle were only the number of connections in a network (G), or the full profile of 16 Motif occurrence (H). \textbf{(I)} Prediction accuracy for test networks, based on oracles that used different sets of network features (shown on x-axis), measured by the $R^2$ value of the True versus Predicted Utility. Each set of features was evaluated across 20 train/test splits, with 50\% of the networks in each dataset. Error bars indicate the standard deviation across the splits.}
    \label{fig:Figure4}    
\end{figure}

We examine directly the computational predictive power of different local network structures for $N=4$ networks. As expected, networks with more connections generally achieve higher Utility, but with substantial variance of Utility for the same density of connections (Fig.~\ref{fig:Figure4}A). Moreover, some sparse networks outperform denser ones, validating the notion that the arrangement of connections, and not just their count, determines computational capability. Networks with more cycles of length 2 or 3 also tend to perform better (Fig.~\ref{fig:Figure4}B-C), whereas ``sinks'' -- neurons with no outgoing connections, which cannot share their computations with the rest of the network, but only with the readout neuron -- impair performance (Fig.~\ref{fig:Figure4}D). We also examined how the presence of each of the 16 non-isomorphic 3-node directed motifs relates to network Utility (Fig.~\ref{fig:Figure4}E; see \nameref{Methods} for the motif decomposition algorithm). For each motif, we aggregate all networks containing that motif and compare their Utility distributions, allowing us to characterize how specific local connectivity patterns shape computational performance. (See Supp. Fig.~\ref{fig:structure_vs_accuracy_fig} for the analogous comparison using Accuracy.)

We next ask how well these structural features predict a network's Utility or Accuracy, either individually, or when combined together -- where in each case we train a one-hidden-layer feed-forward neural network to predict the Utility or Accuracy, using a 50/50 train/test split of networks, repeated across 20 random splits (Fig.~\ref{fig:Figure4}F; see \nameref{Methods}). Using only the connection count yields poor predictions (Fig.~\ref{fig:Figure4}G), whereas using the full motif decomposition achieves near-perfect performance (Fig.~\ref{fig:Figure4}H). Strikingly, combining just three scalar features -- connection count, 2-cycle count, and 3-cycle count ($d=3$) -- captures most of the predictive power of the combined set of motifs or that of the full connectivity matrix (Fig.~\ref{fig:Figure4}I). 

Thus, the computational ability of small networks is largely determined by a small number of local structural statistics, with short recurrent cycles playing a central role. We turn to ask how the structure of larger networks shape their function, and whether the insights we have gained from small networks extend to larger ones. 

\subsection{Adding sparsely connected interneurons rescues the computational performance of large networks}
\begin{figure}[H]
    \centering
    \includegraphics[width=\linewidth]{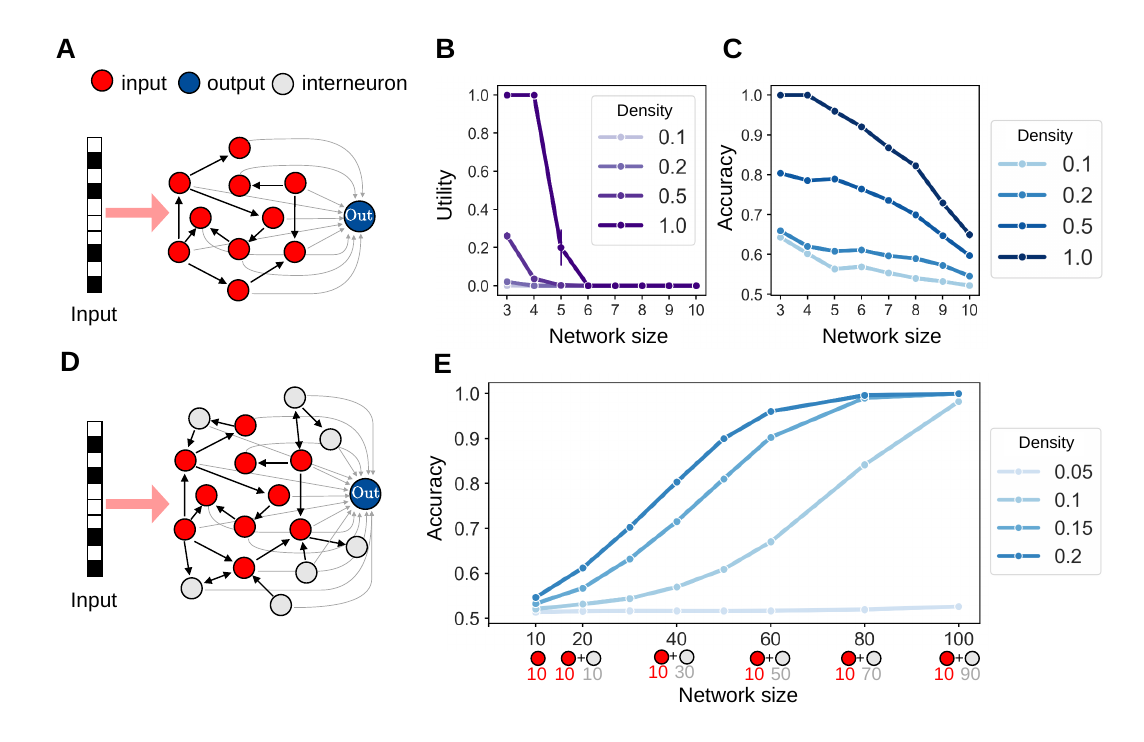}
    \caption{Large networks typically fail to even approximate randomly selected functions, but adding interneurons dramatically improves networks' performance. \textbf{(A)} Schematic of a large network architecture of size $N=10$, composed of input neurons that receive the external input and connect to a readout output neuron, similar to those studied above. \textbf{(B)} Estimated Utility of networks sampled from the Erdős–Rényi model for different values of $N$ and connection density $p$: For every pair of $(N,p)$ values, 100 networks were sampled (except for $N \leq 4$, for which the full space was enumerated) and trained on randomly sampled $N$-bit Boolean functions. Error bars show standard error across samples. \textbf{(C)} Mean approximation accuracy of the same networks and functions. \textbf{(D)} Schematic of network architectures augmented with interneurons that do not receive external input but participate in recurrent computation. \textbf{(E)} Mean approximation accuracy of Erdős–Rényi networks with 10 input neurons and additional interneurons, shown for different values of $N$ and $p$,  where $N$ includes both the input neurons and interneurons; for each configuration, 100 networks were sampled and trained on 100 random 10-bit Boolean functions.
 }
    \label{fig:Figure5}    
\end{figure}

Direct extension of the analysis we have performed on small networks to larger ones is infeasible due to the exponential growth in both the number of directed graphs and the number of Boolean functions, and so we resort to sampling-based approaches for both network architectures and functions: We randomly sample recurrent architectures from the Erdős–Rényi random graph model (see \nameref{Methods}) and train them to compute randomly sampled $N$-bit Boolean functions. Surprisingly, estimating larger networks' computational abilities gives a rather pessimistic picture: While for $N \leq 4$ the fully connected architecture successfully computes essentially all Boolean functions (Fig.~\ref{fig:Figure5}B), performance deteriorates rapidly as $N$ increases. Already for $N=5$ even fully connected networks learn to compute only a small fraction of sampled functions, and for $N\ge6$ performance crashes completely. Networks with sparser connectivity, which we quantify by the fraction of existing synapses out of the $N(N-1)$ possible ones, perform substantially worse. A similar trend appears when we consider the ability of different networks to approximate the functions rather than exact computation: the mean approximation accuracy across sampled networks decreases sharply with network size across all connectivity densities (Fig.~\ref{fig:Figure5}C). 

This rapid decline is consistent with classical results from circuit complexity, which show that most Boolean functions require exponentially large circuits of binary gates to compute \cite{shannon_1949}, and are therefore hard to learn under standard complexity-theoretic assumptions \cite{kearns_cryptographic_1994}. These results reflect that arbitrary Boolean functions may be intrinsically difficult to compute using recurrent networks of a matching size. It could be that this is an inherent limitation of aiming at the total set of Boolean functions, and that different network architectures ``cover'' different parts of the function space. Moreover, biological neural networks obviously rely on much larger architectures, and in some cases circuit design seems to be optimized for specialized designated computations. On the other hand, other neural circuits, such as cortical columns are often considered to be highly adaptable and capable of implementing very different functions. 

Inspired by real neural network architectures, we find that a simple architectural expansion dramatically changes the ability of our networks and enables them to overcome these seemingly hard classes of functions. We augment the networks with additional ``interneurons'', namely neurons that do not directly receive external input, but participate in the recurrent dynamics (see Fig.~\ref{fig:Figure5}A and Fig.~\ref{fig:Figure5}D for an illustration of networks with or without interneurons). The resulting architectures exhibit a striking increase in computational performance. As the number of interneurons grows, the approximation accuracy of randomly sampled recurrent networks improves rapidly (Fig.~\ref{fig:Figure5}E). Critically, this is the case even for very sparsely connected networks (that were useless without the interneurons). Thus, with additional  interneurons, even sparse networks become capable of learning to compute arbitrarily complex Boolean functions. This is particularly interesting since these networks lack any imposed architectural structure, as all networks are sampled from the Erdős–Rényi model.

\begin{figure}[H]
    \centering
    \includegraphics[width=\linewidth]{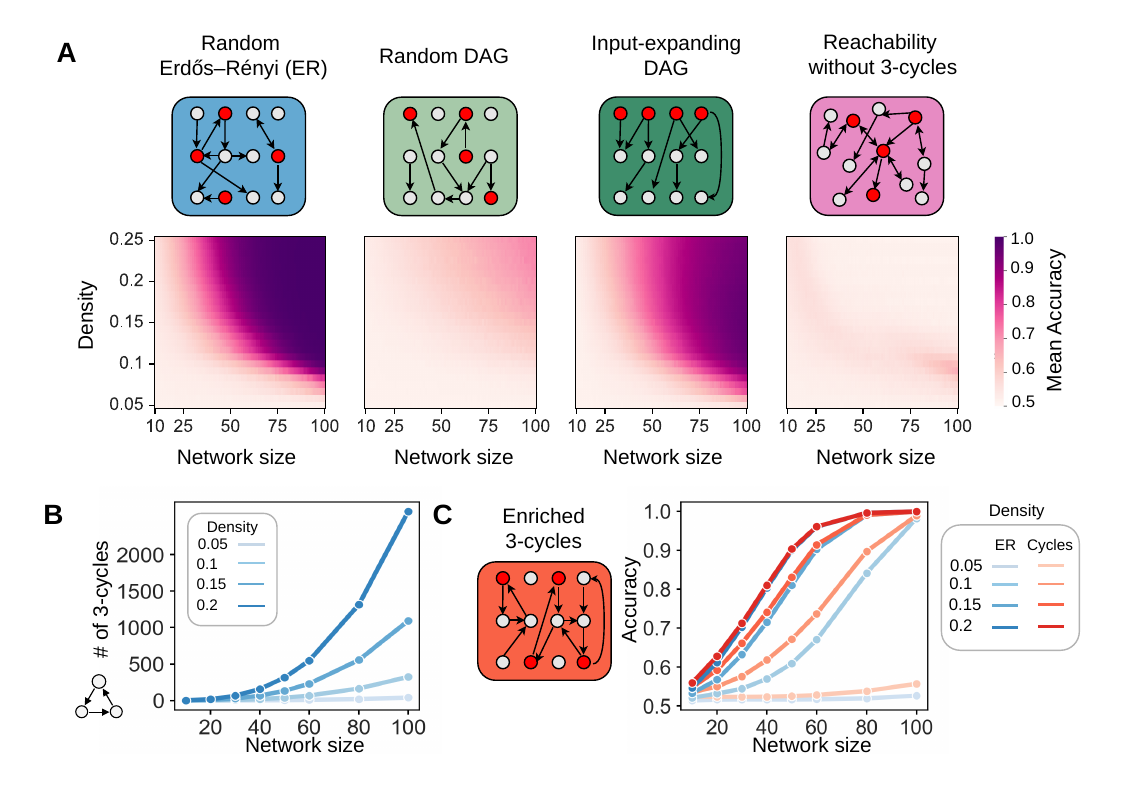}
    \caption{
    \textbf{Structured random graph ensembles reveal a functional role for short cycles.}
    \textbf{(A)} Mean accuracy for networks sampled from four graph models: Erdős–Rényi (ER), Directed Acyclic Graphs (DAG), a structured acyclic model with propagating inputs (Input-expanding DAGs), and reachability-enhanced networks without 3-cycles. Networks were sampled with sizes $N \in [10,100]$ and densities $p \in [0.05,0.25]$, where all networks have exactly 10 input neurons. Each cell in the heatmaps is the mean accuracy of 100 sampled networks with these $N,p$ values trained on the same 100 random 10-bit Boolean functions (\nameref{Methods}). \textbf{(B)} Abundance of 3-cycles in ER networks as a function of network size, shown for several connection densities, averaged over 100 networks per $(N,p)$ configuration. \textbf{(C)} Left: schematic of the ``Enriched 3-Cycles'' ensemble (ER graphs enriched with additional short cycles; see \nameref{Methods}). Right: mean accuracy of ER networks (blue lines) is shown together with the mean accuracy of Cycles-enriched networks (red lines) as a function of network size; Lines of different hues show different connection densities; in all cases networks enriched with cycles outperform the ER ones, for corresponding density values. Statistical significance of cycles-enriched vs. ER ensembles was assessed per $(N, p)$ configuration with a one-sided Mann--Whitney $U$ test, Bonferroni-corrected. Cycles-enriched networks significantly outperformed ER in all comparisons ($p < 0.05$), except for $N\geq 80$ and $p\geq 0.15$ where ER networks already contain an abundance of short cycles (Fig.~\ref{fig:Figure6}B).
    }
    \label{fig:Figure6}    
\end{figure}

We then asked whether imposing additional structural features to the connectivity could further improve networks' capacity. We therefore tested different classes of networks with interneurons, and compared their mean approximation accuracy as a function of network size and connection density (Fig.~\ref{fig:Figure6}A). The Erdős–Rényi (ER) networks (corresponding to those in Fig.~\ref{fig:Figure5}) improve with both size and density, yielding a strong accuracy gradient over the $(N,p)$ plane. Notably, and following the predictive nature of cycles in small networks -- these ER graphs contain many short cycles, and in particular 3-cycles, whose expected abundance increases rapidly with network size (Fig.~\ref{fig:Figure6}B). We therefore asked whether cycles are a key contributor to network performance. To test this, we studied an ensemble of Directed Acyclic Graphs (DAGs) with matched connection density $p$ for each $N$ value and found that they perform poorly across the full range of sizes and densities, suggesting that removing cyclic recurrence severely limits computation. We then asked whether this failure is purely due to the lack of cycles, or perhaps cycles were contributing a different structural property, such as an improved propagation of information through the network. To isolate this factor, we introduced a structured acyclic ensemble (``Input-expanding DAG'') in which input neurons are forced to project broadly to downstream interneurons (see \nameref{Methods}). Input-expanding DAGs perform substantially better than unconstrained DAGs, indicating that deliberately ensuring input propagation can partially rescue performance even without cycles, but they remain less effective than ER networks over much of the parameter range. We then asked whether the advantage of cyclic architectures can be explained solely by improved signal propagation. Because our recurrent dynamics run for $T=3$ steps, information can traverse at most three ``hops'' during computation. We therefore construct a ``Reachability-without-Cycles'' ensemble of networks that explicitly maximizes \emph{3-step reachability}, i.e. the number of neurons reachable from each input within three directed hops, while prohibiting 3-cycles (\nameref{Methods}). If cycles were merely a proxy for efficient propagation, then these networks should be on par with ER networks at matched $(N,p)$. Instead, networks with high reachability but not 3-cycles perform poorly across all sizes and densities, often worse even than standard DAGs. 

We therefore asked whether directly enriching ER networks with more cycles, thus creating the ``Enriched 3-cycles'' ensemble (\nameref{Methods}) would improve performance even further. We find that this is indeed the case, consistent with the interpretation of the other classes of networks (Fig.~\ref{fig:Figure6}C). Moreover, we find that the improvement is most dominant in the sparse regimes, whereas it diminishes at higher densities where ER networks already contain many short cycles.

\section{Discussion}
By systematically training hundreds of millions of pairs of recurrent neural networks and target functions, we mapped how network connectivity shapes the computational abilities of finite recurrent neural networks. We found that most Boolean functions are intrinsically difficult to compute, and that most recurrent neural network architectures can learn to compute only a small subset of them. While increasing connectivity generally improves performance, we find very high variance in the computational ability of networks with similar connection density. We showed that short recurrent cycles of length 2 and 3 are key factors in shaping the computational capacity and are dominant in the minimal architectures required to compute particular functions, whereas acyclic organization and sink nodes impair performance. For small networks we showed that their computational performance can be predicted accurately from the connection count and short-cycle counts. Extending our analysis for more than a handful of neurons showed that typical networks fail to even approximate typical functions. But, inspired by interneurons in biological neural networks, we found that adding a sparsely connected and small set of such intermediate neurons changes the capacity of larger networks dramatically, making them highly capable, and that here too, short cycles both predicted and improved computational capacity. 

While consistent with the computational interpretations of small network motifs \cite{alon_milo_motifs}, the importance of short cycles here is notable, given that many of the dominantly studied and used artificial neural networks are acyclic circuits -- from multilayered perceptrons \cite{rumelhart_learning_1986}, through convolutional networks \cite{lecun_gradient-based_1998}, to transformers \cite{vaswani_transformers}. This is particularly interesting since in our finite and resource-limited networks, which echo the constraints of many biological circuits, we find that local cycles are often the minimal architectures capable of computing many functions. One natural interpretation is that cycles provide a form of short-term memory, allowing information to persist in the network and be re-processed across the network's runtime rather than dissipating after a few feedforward steps. This notion is supported by the failure of our reachability-matched acyclic model, which propagates information but cannot retain it. 

Future work would explore even larger networks, and aim at uncovering new and more refined principles of the architecture, beyond the local ones we have found here, which govern high-capacity connectivity patterns, especially for the networks with sparsely connected interneurons. Another extension would be to go beyond the two classes of neuronal nonlinearity that we considered here, as they could prove to be qualitatively different. In addition, we have focused here on the case of $T=3$, as it is the minimal time scale that allows for complex dependencies in processing the inputs as well as internal memory in the network, but larger values of $T$ could reveal the potential capacity and limitations of longer recurrent dynamics. We also note that while our choice of Boolean functions is grounded in their expressive power and computational tractability, other classes of functions may reveal more intricate or interesting organization of the space of networks, or the space of functions (from the point of view of networks’ performance). Our analysis has focused on finding a set of network weights that yields the correct computation or its approximation, and ignored variability in learning success due to our learning paradigm or our sampling of initial conditions. Quantifying how reliably networks may reach a working solution, across random initializations of the networks, will give another dimension of network design – distinguishing between network topologies that are sensitive to their initial conditions, implying a rugged landscape of the space of networks that share the same topology with different connection weights vs. topologies that are so effective that they are insensitive to the initial conditions \cite{marder_prinz}. We also note that other learning mechanisms could reveal more capable network designs, as well as exploring wider range of initial conditions for training networks, or alternatively evolve architectures under selection for computational performance \cite{arcas_computational_2024}, and ask which dominant structures emerge.

We hope that the analysis we presented here would benefit the study of connectomes of real neural circuits and their function, and maybe also serve as a framework for exploring architectural design of artificial ones. In particular, the framework we presented here would allow to ask which classes of architectures would be beneficial for specific classes of computational tasks, namely to characterize the internal structure of our catalog and accuracy matrices for large and diverse classes of networks. The capacity of fixed randomly connected recurrent networks to supply rich dynamics while only a readout or a feedback input is trained \cite{maass_real-time_2002,sussillo_generating_2009, jaeger_harnessing_2004}, offer another potential extension of our work, namely asking whether building structure into these ``reservoirs'' could improve their performance or construction. On a more theoretical direction, future work would aim at connecting our findings with analytic approaches that derive network dynamics directly from connectivity, such as the graph-theoretic rules for threshold-linear networks that show how specific motifs give rise to particular dynamical regimes \cite{curto_graph_rules, curto_graphical_2025}. Establishing similarly rigorous links between connectivity motifs and the computations a network can perform would be instrumental in bridging network architecture, dynamics, and function.

\section*{Acknowledgements}
We thank Or Samimi Golan for his work on an early version of this project, Oren Richter, Roy Urbach, and other members of Schneidman’s lab for critical suggestions and insights, as well as Itay Talpir. This work was supported by Simons Collaboration on the Global Brain grant 542997, Israel Science Foundation grant 137628, Azrieli Institute for Brain and Neural Sciences and the Hedda, Alberto, and David Milman Baron Center for Research on the Development of Neural Networks of the Weizmann Institute, as well as the Knell Family Institute for Artificial Intelligence, Martin Kushner Schnur, and Mr. \& Mrs. Lawrence Feis. ES is the incumbent of the Joseph and Bessie Feinberg Chair.

\newpage
\section*{Methods}\label{Methods}
\subsection*{Modeling recurrent networks of neurons}
Our recurrent neural networks consist of $N$ neurons, where the connectivity between them is defined by a directed graph $G = (V,E)$ with $|V| = N$. We denote the connection weight from neuron $i$ to neuron $j$ by $W_{ij} \in \mathbb{R}$, and self-connections were excluded (i.e., $\forall i: W_{ii}=0$).
Neural activity of the neurons at discrete time $t$ is given by ${\bf x}(t) \in \mathbb{R}^N$. Given the Boolean input ${\bf x}_0 \in \{0,1\}^N$, the network is initialized as ${\bf x}(t=0) \ = \ {\bf x}_0$.
\\\\
For $t = 1, \dots, T$, neuronal states evolve according to 
\begin{align*}
    x_j(t+1) = \sigma ( \sum_{i \ne j } W_{ij} \cdot x_i(t) + b_j  )
\end{align*}
where $b_j \in \mathbb{R}$ is a bias term and $\sigma(\cdot)$ is a nonlinear activation function. We examined both the Sigmoid and the ReLU activation functions, and found similar learning performance (see Supplementary Material). 
\\\\
After $T$ time steps, the network output is computed by a linear readout neuron:
\begin{align*}
    x_{\text{out}} = \phi(\sum_{i=1}^N v_i x_i (T)  + b_\text{out})
\end{align*}
where $v_i \in \mathbb{R}$ are readout weights,  $b_{\text{out}} \in \mathbb{R}$ is a bias term and $\phi(z) = \mathbbm{1}[z \ge 0]$ is the Heaviside threshold function and $\mathbbm{1}[\cdot]$ is the indicator function, equal to $1$ when its argument is true and $0$ otherwise. Thus, the full set of trainable parameters is $\theta = \{ W, b, v, b_{\text{out}} \}$.

\noindent All simulations and training procedures were implemented in PyTorch~\cite{pytorch}.

\subsection*{Training networks with Backpropagation-Through-Time}
A network ``topology'' is defined by the unweighted directed graph $G = (V,E)$. Training the network to compute a given function is done by optimizing the weights of the connections in $E$, i.e., no new connections are introduced, and so if $(i, j)\notin E$ then $W_{ij}=0$.

Given a network–function pair $(\mathcal{N}, f)$ we optimize the network parameters $\theta = \{ W, b, v, b_{\text{out}} \}$ so that the network output approximates $f(x)$ for all $x \in \{0,1\}^N$. In each epoch, the network is trained on the full truth table $\{(x, f(x)) : x \in \{0,1\}^N\}$, and parameters are updated to minimize the Binary Cross-Entropy (BCE) loss computed over all $2^N$ input patterns:
\begin{align*}
    \mathcal{L}_{\text{BCE}} = -\frac{1}{2^N}\sum_{x \in \{0,1\}^N} \left[f(x)\log \hat{y} + (1-f(x))\log\big(1 - \hat{y}\big) \right]
\end{align*}
where $\hat{y} = \sigma(\sum_{i=1}^N v_i x_i (T)  + b_\text{out})$ and $\sigma$ is the Sigmoid function, meaning that we optimize the network based on the readout neurons' output prior to thresholding. Training is performed for 500 epochs using Backpropagation-Through-Time (BPTT)~\cite{BPTT} to propagate gradients across the recurrent unrolled dynamics, and parameters are optimized using the Adam optimizer~\cite{adam_optimizer}. To account for variability due to random initializations of weights, each network–function pair is trained independently across 10 random initializations.  Initial weights and biases are sampled independently from a normal distribution $\mathcal{N}(0,1)$. We use early-stopping if the network learned to accurately compute the function, i.e. reaches 100\% classification accuracy on all $2^N$ inputs. If the network did not learn to compute the function, we record the best-performing run across all initializations.

After training each of the network-function pairs, we record in the Catalog Matrix $\mathcal{C}$ whether network $\mathcal{N}_i$ successfully learned to compute function $f_j$ in at least one of the initializations. $\mathcal{C}_{ij}=1$ indicates that network $\mathcal{N}_i$ produces the correct output for all $2^N$ inputs of $f_j$:
\begin{align*}
    \mathcal{C}_{ij} = \prod_{x\in \{0,1\}^N} \mathbbm{1}[\mathcal{N}_i(x) = f_j(x)]
\end{align*}
where $\mathbbm{1}[\cdot]$ is the indicator function.

Similarly, the Approximation Matrix $\mathcal{A}$ denotes the highest accuracy network $\mathcal{N}_i$ achieves on function $f_j$ over the different initializations:
\begin{align*}
    \mathcal{A}_{ij} = \frac{1}{2^N}\sum_{x\in \{0,1\}^N} \mathbbm{1}[\mathcal{N}_i(x) = f_j(x)]
\end{align*}

\subsection*{Motif decomposition}
For a directed graph $G = (V,E)$ with nodes $V$ ($|V| = N$) and edges $|E|$, a motif of size $k$ is defined as an induced subgraph $G' = (V',E')$ with $V' \subseteq V$, $|V'| = k$, and $E'$ containing all edges of $E$ between nodes in $V'$. We focus here on motifs of size $k=3$, where there are 16 non-isomorphic directed motifs. For a graph with $N$ nodes, there are ${N \choose 3}$ distinct induced subgraphs of size three. While motif enumeration can be computationally demanding in large networks~\cite{motif_decomp}, the small size of our networks permits exhaustive enumeration. Thus, for each graph $G$, we iterate over all ${N \choose 3}$ node triplets, classify each induced subgraph into one of the 16 triad types, and obtain a motif census vector $c(G) \in \mathbb{N}^{16}$, 
where $c_i(G)$ denotes the number of occurrences of motif type $i$ in $G$.

\subsection*{Predicting computational ability from network structure}
To assess whether structural properties predict computational performance, we trained feed-forward multilayer perceptrons (MLP) to map structural features of a network to a network's \textit{Utility} or \textit{Accuracy}. Thus, for each network $\mathcal{N}$, we use the following structural features $\psi(\mathcal{N})$: (1) Number of connections, (2) Number of 2-cycles, (3) Number of 3-cycles, (4) Binary connectivity matrix, flattened and without the diagonal, which is always set to $W_{ii}=0$, (5) Motif count vector $c \in \mathbb{N}^{16}$.

We evaluate the predictive power of each feature individually, as well as all possible feature combinations. For a chosen feature map $\psi$, we construct a dataset $\mathcal{D} = \{(\psi(\mathcal{N}), y(\mathcal{N})) \}$ from all networks, where $y(\mathcal{N})$ denotes either Utility or Accuracy of network $\mathcal{N}$. The MLP is trained over random splits of the set of networks and Utility or Accuracy, with 50\% used for training and 50\% for testing; To obtain stable performance estimates, we repeat the random train/test split 20 times and report the averaged test performance. We train a fully connected network with a single hidden layer and ReLU activations to minimize the Mean Squared Error $ \mathcal{L}_{\text{MSE}} = \frac{1}{m} \sum_{i=1}^{m} \left(y_i - \hat{y}_i\right)^2$, with the optimization performed using Adam for 500 epochs.

\subsection*{Sampling networks with $N\geq5$ neurons}
For networks with $N \ge 5$, exhaustive enumeration of all $2^{N(N-1)}$ possible directed graph topologies becomes challenging, and we therefore rely on sampling. To explore how structural properties influence performance, we generate networks using different random graph models, each inducing distinct structural characteristics:

\noindent {\bf Erdős–Rényi model.} The Erdős–Rényi model \cite{erdos1960} is a canonical model for generating random graphs, where each ordered pair of distinct nodes is connected independently with probability $p$. In the directed graph version that we use here, the graph $G(N,p)$ is a randomly directed graph with expected edge density $p$.

\noindent {\bf Directed Acyclic Graph (DAG) model.} We construct Directed Acyclic Graphs (DAGs) with a prescribed edge density $p$: We first pick an arbitrary ordering of the nodes and add a directed edge from every node to all nodes that come after it in the ordering. Because every edge points ``forward" along the ordering, no directed cycle can form.  The resulting graph contains $\frac{N(N-1)}{2}$ edges, which is the maximum possible for an acyclic graph on $N$ nodes. The connectivity matrix $W$ is strictly upper-triangular under this ordering. To achieve the desired density $p$, we prune edges at random until the total number of edges reaches $\lfloor p \cdot N(N-1) \rfloor$. Since edges are only removed, the acyclic property is preserved throughout the process.

\noindent {\bf Input-expanding DAG model.} Randomly pruning the edges in the DAG model might yield graphs in which the input neurons are barely connected to the rest of the network. The Input-expanding DAG are ones where we first prune connections that do not originate from an input node, and only when there is no other choice we prune input-related edges. This yields networks that are specifically engineered to have their input expand to the rest of the network, and are still sparse DAGs.

\noindent {\bf Reachability without 3-cycles model.} We introduce a sampling procedure that promotes long-range reachability while explicitly eliminating cycles of length 3. We recall that for a directed graph $G$ with connectivity matrix $W$, the $(i,j)$ entry of the matrix  $W^k$ counts the number of directed paths of length exactly $k$ from node $i$ to node $j$. Thus, $(W + W^2 + W^3)_{ij} > 0$ if there exists a directed path from $i$ to $j$ of length at most three. We then define the \emph{3-reachability} of node $j$ as the number of nodes from which $j$ can be reached via a directed path consisting of no more than three edges:
$$
r(j) = \bigm| \{ i \ne j : (W + W^2 + W^3)_{ij} > 0\}\bigm|
$$
where $|\cdot|$ denotes the cardinality of the set. The 3-reachability of a network is given by the mean over nodes, $R(G) = \frac{1}{N} \sum_i r(i)$, and quantifies how broadly information spreads through the network within three time steps. 
Thus, after sampling a directed Erdős–Rényi graph $G(N,p)$, we iteratively remove directed 3-cycles while preserving the total number of edges: at each iteration we pick a 3-cycle, remove one of its three edges, and add a new edge between an unconnected pair of nodes. The added edge is chosen to maximize $R(G)$, which we do by going over all candidate non-edges $i \rightarrow j$, computing $R(G)$ that would result from adding it, and select the pair giving the highest value. If no candidate non-edge increases $R(G)$, a random non-edge is added instead. This process continues until no 3-cycles remain. The resulting network retains the same edge density as the original $G(N,p)$ graph, but contains no cycles of length 3 and exhibits enhanced reachability.

\noindent {\bf Enriched 3-cycles model.} To investigate the specific role of short recurrent motifs, we introduce a sampling procedure that generates random directed networks with edge probability $p$, while explicitly enriching the abundance of length-3 cycles. The algorithm first determines the total number of edges $n_e \sim \text{Binomial}(N(N-1), p)$, matching the expected density of an Erdős–Rényi graph. Edges are then added iteratively with a bias toward forming directed 3-cycles. Starting from $N$ unconnected nodes, at each iteration, we add a connection aiming to complete a full 3-cycle among three randomly selected nodes. If there is an existing directed path of length two, $i \rightarrow k \rightarrow j$, we close the cycle by adding the edge $j \rightarrow i$. If no such path exists among all possible triplets, we select an existing edge $i \rightarrow j$ and introduces a third node $k$, adding the edges $j \rightarrow k$ and $k \rightarrow i$ whenever possible to form a new 3-cycle. This iterative procedure continues until the prescribed number of edges is reached. The resulting graph has approximately edge density $p$, but contains a significantly higher number of length-3 cycles compared to a standard $G(N,p)$ random graph.

\subsection*{Sampling Boolean functions}
For functions over $N \geq 5$, exhaustive enumeration of all $2^{2^N}$ Boolean functions becomes impossible, and so we rely on sampling. We sample Boolean functions by randomly and independently assigning each of the $2^N$ binary output values a value of either 0 or 1 with probability $p=0.5$, yielding uniformly random truth tables. For the case of $N=5$, we also explored a structured sampling scheme designed to span a wide range of computational difficulty by matching the empirical joint distribution of Fourier degree and Total Influence over the full function space (see Supplementary Fig. \ref{fig:sampling_comparison}): This scheme produced network performance distributions that were very similar to those obtained with uniform sampling, and so we used uniform sampling throughout the main text.

\newpage 
\section*{Supplementary Material}
\setcounter{figure}{0}
\renewcommand*{\thefigure}{S\arabic{figure}}

\subsection*{Low dimensional structures of the catalog matrices}
\begin{figure}[H]
    \centering
    \includegraphics[width=\linewidth]{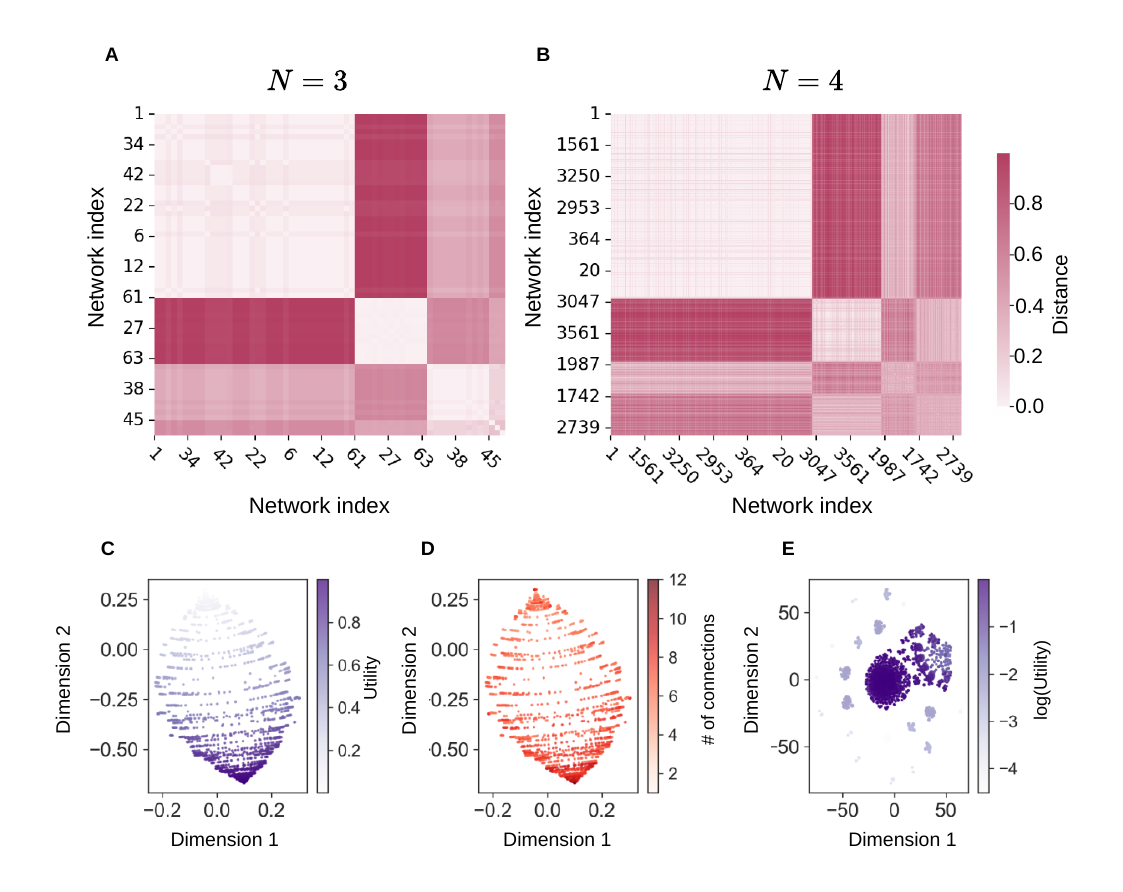}
    \caption{{\bf Clustering networks based on the similarity of their functional capabilities.} (\textbf{A},\textbf{B}) Distance matrices for networks of 3 and 4 neurons . The distance between a pair of networks $(\mathcal{N}_i, \mathcal{N}_j)$ is the normalized Hamming distance between their functionality vector, i.e. the binary vector denoting which of the $2^{2^N}$ functions they fully compute (Equation~\ref{eqn:Eq5}). (\textbf{C},\textbf{D}) Two-dimensional embedding of the space of networks using Multidimensional Scaling (MDS) based on the functional Hamming distance. Each point represents a network, colored by Utility (C) or number of connections (D). (\textbf{E}) Two-dimensional embedding of networks using Principal Component Analysis (PCA), colored by log-Utility.}
    \label{fig:SuppFig_CatMat_Distances}    
\end{figure}

Using the Catalog and Approximation matrices, we can define a distance metric between pairs of networks. We measure the similarity between network pairs based on their functional capabilities, comparing the set of functions they can solve. Each network is described by the binary vector of length $2^{2^N}$, where each entry indicates whether the network solves a particular function. The similarity between two networks is then given by the normalized Hamming distance between their vectors: 
\begin{align}
    \text{dist}(\mathcal{N}_i, \ \mathcal{N}_j) = 1 - \frac{1}{2^{2^N}} \sum_{f \in \mathcal{F}}  \mathbbm{1}[ \mathcal{C}_{if} = \mathcal{C}_{jf} ]
\label{eqn:Eq5}   
\end{align}
where $\mathbbm{1}[\cdot]$ is the indicator function.
The distance matrices for networks with 3 neurons and networks with 4 neurons are shown in Figure~\ref{fig:SuppFig_CatMat_Distances}A,B. In both cases, the networks can be grouped into a few distinct clusters. For $N=3$, we identify three well-defined clusters with little variability within each, suggesting that many networks exhibit similar behaviors. However, for $N=4$, the clustering becomes less distinct, with greater variability within each group. This suggests that perhaps there is no  simple organizing principle for defining the capabilities of a network. To further probe whether there is an underlying organization in the space of networks, we explored low-dimensional representations of their functional similarities. We first applied Multidimensional Scaling (MDS) \cite{cox_multidimensional_2008} to the pairwise distance matrix defined above. Figures~\ref{fig:SuppFig_CatMat_Distances}C,D show a two-dimensional MDS embedding, with networks colored by Utility and by connection count, respectively. When colored by Utility (Fig.~\ref{fig:SuppFig_CatMat_Distances}C), the embedding reveals a clear gradient, with low-Utility networks separated from high-Utility ones. In contrast, coloring the same embedding by connection count (Fig.~\ref{fig:SuppFig_CatMat_Distances}D) shows no comparable structure: networks with similar density can occupy widely separated regions and exhibit vastly different Utilities. This indicates that the number of network connections alone does not explain the functional organization revealed by the embedding. We complement this analysis with Principal Component Analysis (PCA) applied directly to the functional representations of networks. The resulting two-dimensional PCA embedding (Fig.~\ref{fig:SuppFig_CatMat_Distances}E) does not reveal a clear low-dimensional organization based on functional capabilities, reinforcing the notion that the space of network functions is not easily captured by simple linear projections.

\subsection*{Catalog matrices of ReLU and sigmoid activations show similar networks' capacity}
\begin{figure}[H]
    \centering
    \includegraphics[width=0.35\textwidth]{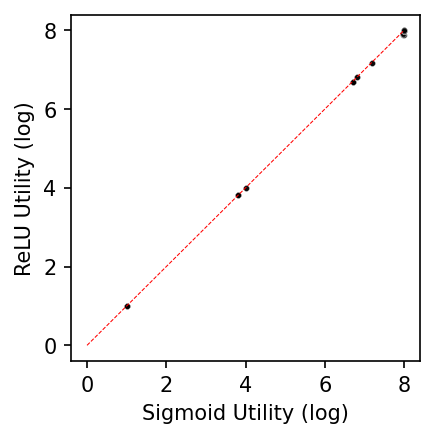}
    \caption{{\bf Network performance is almost identical between different activation functions.} Utility of $N=3$ networks, trained with either ReLU or Sigmoid. Each dot is a network. Almost all networks achieve the same Utility for both activation functions.}
    \label{fig:suppfig_relu_vs_sigmoid}    
\end{figure}

We studied the space of networks with neurons that use either a ReLU activation function $\sigma$, where \textbf{ReLU} : $\sigma(z) = \max(0, z)$ or Sigmoid where $\sigma(z) = \frac{1}{1+e^{-z}}$. While in the main text we showed the results for the Sigmoid case, supplementary Figure~\ref{fig:suppfig_relu_vs_sigmoid} compares the Utility of all 64 $N=3$ networks on all 256 3-bit Boolean functions, trained with either ReLU or Sigmoid activations. The Catalog matrices generated with these activation functions showed 99.4\% agreement, indicating that they behave very similarly.

\subsection*{Adding self-connections for neurons improves networks' capacity}
\begin{figure}[H]
    \centering
    \includegraphics[width=0.95\textwidth]{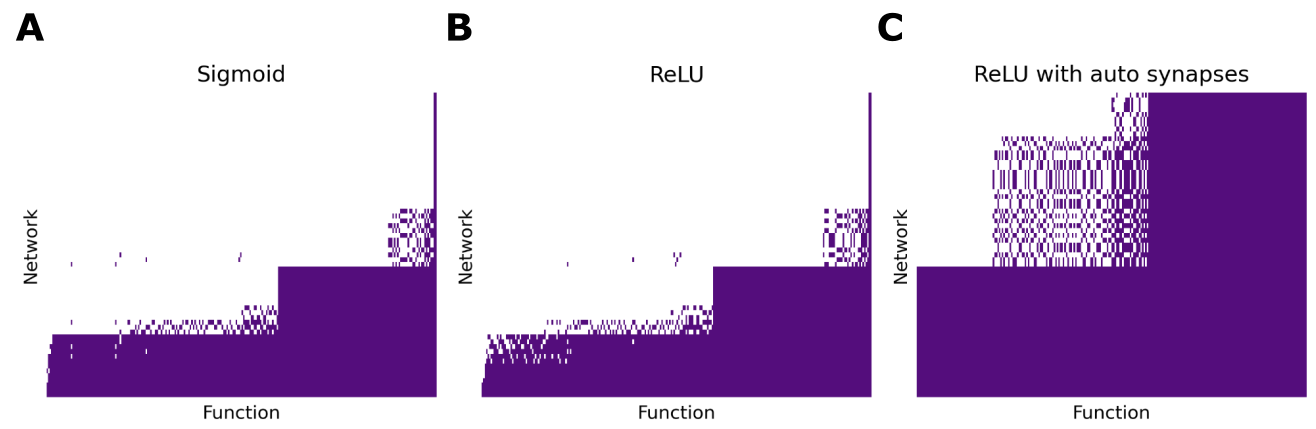}
    \caption{{\bf Catalog matrices comparing the Utility of networks with either Sigmoid or ReLU activations, and networks with self-connections}. Every Catalog Matrix consists of 64 3-neuron networks, and all 256 3-bit Boolean functions. The performance of ReLU and Sigmoid networks was almost identical: they showed 99.4\% agreement in their Catalog Matrix entries. Networks with self-connections exhibit a noticeable improvement in performance.}
    \label{fig:catmats_sigmoid_relu_autosynapses}    
\end{figure}

\begin{figure}[H]
    \centering
    \includegraphics[width=\textwidth]{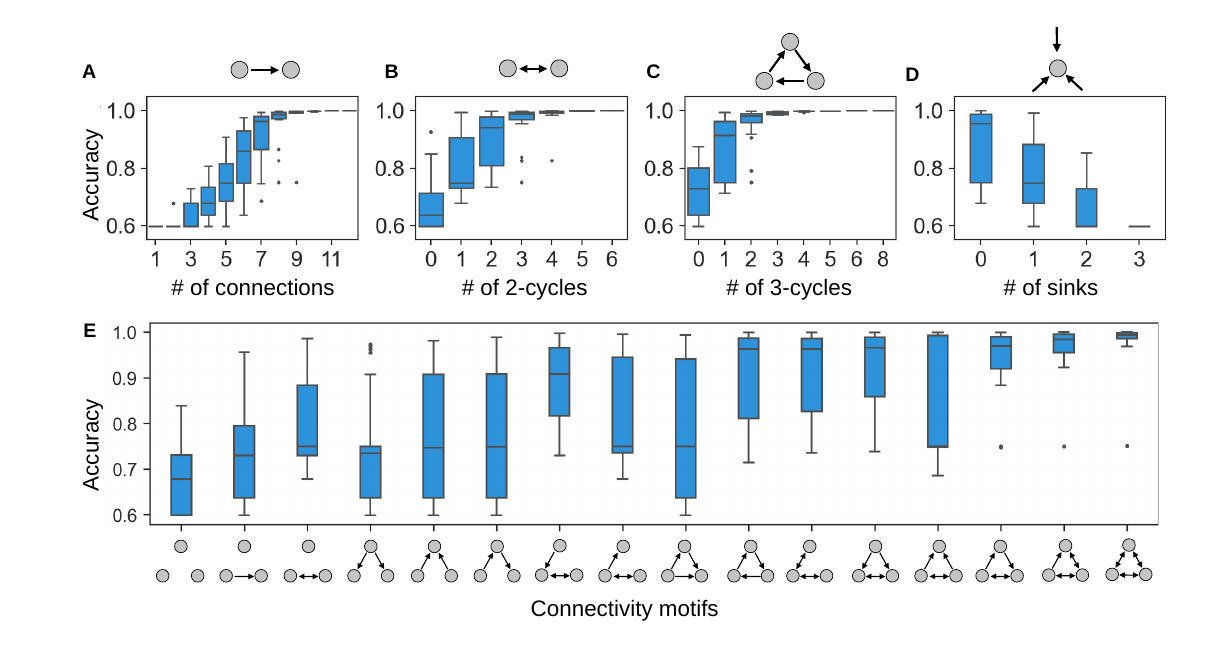}
    \caption{{\bf Network Accuracy as a function of structural properties} \textbf{(A-D)} The effect of number of connections, 2-cycles, 3-cycles and sinks (respectively) on Accuracy of $N=4$ networks across all 4-bit Boolean functions. \textbf{(E)} Networks are decomposed into motifs of size 3. For every possible motif, we record the performance distribution of all networks containing that motif. }
    \label{fig:structure_vs_accuracy_fig}    
\end{figure}

Throughout the main text our networks had no self-connections, meaning that for all neurons $W_{ii} = 0$. However, we also mapped a catalog matrix for $N=3$ where networks have self-connections, to explore their potential benefits. As expected, such networks showed improved  performance, as shown in Supp. Figure \ref{fig:catmats_sigmoid_relu_autosynapses}.

%\newpage
\subsection*{Relation of different structural properties of networks to their Accuracy}
We examine how the structural features analyzed in Figure~\ref{fig:Figure4} relate to the Accuracy of $N=4$ networks. The resulting trends closely mirror those observed for Utility.

\newpage
\subsection*{The effect of runtime on network performance}
\label{sec:runtime}
In our model, network dynamics are governed by Equations \ref{eqn:Eq1} and \ref{eqn:Eq2} and the neuronal states are updated for $T$ steps, and the computation's output is given at time $T$. Thus, $T$ plays a key role in network dynamics, enabling information to persist over time, effectively creating a form of  short term memory. Additionally, runtime influences the synchronization of computations across networks. For example, if a network requires $T$ timesteps to compute a function, downstream networks must adapt to the timing of upstream computations. In the main text we focused on runtime of $T=3$, and here we present preliminary results for how networks behave for other runtimes.

\begin{figure}[H]
\centering \includegraphics[width=\textwidth]{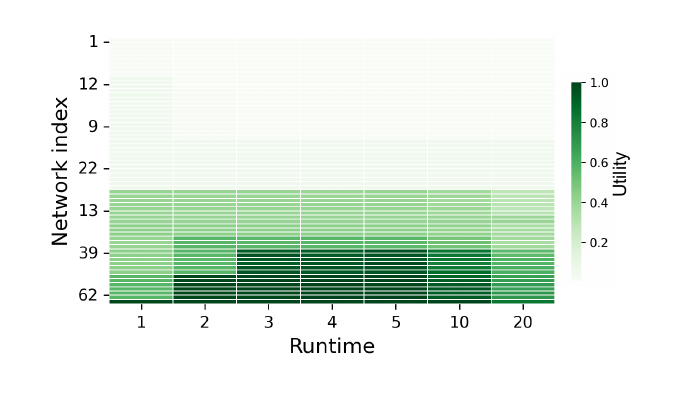}
    \caption{{\bf The effect of runtime on $N=3$ networks.} All 64 networks were trained to solve all 256 3-bit Boolean functions, for different runtimes. We record the Utility of each network, i.e. the fraction of functions it successfully solves. Heatmap rows are sorted by the mean Utility of each network.}
    \label{fig:runtime}    
\end{figure}

We therefore trained all 64 $N=3$ networks on all 256 3-bit Boolean functions, testing runtimes $T \in \{1, 2, 3, 4, 5, 10, 20\}$. Figure~\ref{fig:runtime} shows the Utility achieved by every network for the different runtime values. We observe peak performance at runtimes of $T\in \{3,4,5\}$, with performance deteriorating for both shorter and longer runtimes. This decline may be a training artifact of gradient descent -- factors like vanishing gradients might make training harder at higher runtimes. Moreover, it is unclear whether a network $\mathcal{N}_i$ that learns to solve a function $f_j$ in $T$ steps can always find a solution using runtime of $T_1$ steps, with $T_1 > T$. It is unclear whether it is always possible to ``stretch" or ``delay" a computation by changing the value of $T$, which can be important for synchronizing different networks. These questions should be at the core of a future analysis of the runtimes of recurrent networks.

\subsection*{Structured sampling of Boolean functions for $N=5$}
The uniform Bernoulli sampling of functions that we used for $N \ge 5$ produced functions that were hard for the networks that we studied. However, since Boolean function analysis and circuit complexity have established that almost all such functions require circuits of exponential size to compute \cite{shannon_1949, odonnelBoolFunctions}, we asked whether a biased sampling that would aim at sampling functions by their complexity would give different results. 

We therefore quantified the structural properties of Boolean functions using two measures \cite{odonnelBoolFunctions}: First, we recall that every Boolean function $f$ can be written uniquely as a sum over all subsets of input variables, known as its Fourier expansion:
\begin{align*}
    f(x)=\sum_{S\subseteq [N]} \hat{f}(S)\,x^S,
\end{align*}
where the sum ranges over all $2^N$ subsets $S$ of the index set $[N] = \{1, 2, \ldots, N\}$, and $x^S = \prod_{i\in S} x_i$ is the product of the input variables indexed by $S$. Thus, $x^S$ equals the parity of the bits in $S$. The Fourier coefficient $\hat{f}(S) \in \mathbb{R}$ is given by $\hat{f}(S) = \mathbb{E}_x[f(x)\, x^S]$, with the expectation taken over $x$ drawn uniformly from $\{-1,1\}^N$, and quantifies how strongly $f$ correlates with the parity over the variables in $S$. Coefficients with large absolute value indicate that $f$ depends substantially on the joint behavior of those variables, while coefficients close to zero indicate that the corresponding parity contributes little to $f$. The Fourier expansion naturally leads to the definition of a function's degree:
\begin{align*}
    \text{deg}(f) = \max \{ |S| : \hat{f}(S) \ne 0\}.
\end{align*}

\noindent and since low degree functions depend only on low order combinations of bits, this is a structural measure of the complexity of $f$. However, even for functions of the same degree, computational difficulty can vary substantially: solving the AND function on $N$ bits is significantly easier than solving the XOR function on $N$ bits. To distinguish between such cases, we measure the influence of the $i$-th bit on the function's output: The $i$-th bit is pivotal for $f$ if flipping it changes the function's output, i.e. $f(x) \ne f(x^{\oplus i})$ where $x^{\oplus i}$ denotes $x$ with the $i$-th bit flipped. Averaging over all possible inputs gives the influence of the $i$-th bit:
\begin{align*}
    \textbf{Inf}_i[f] = \frac{1}{2^N}\sum_{x\in \{ -1, 1\}^N } \mathbbm{1}[f(x) \ne f(x^{\oplus i})],
\end{align*}
where $\mathbbm{1}[\cdot]$ is the indicator function. The Total Influence $I$ of $f$, measures the expected number of bits whose individual flips change $f$'s output,   
\begin{align*}
    \textbf{I}[f] = \sum_{i=1}^N \textbf{Inf}_i [f],
\end{align*}
\noindent Thus, functions with high total influence are sensitive to input perturbations, whereas functions with low total influence remain stable under most single-bit changes.

Together, Fourier degree and Total Influence provide proxies for the computational complexity of a Boolean function $f$. For $N=5$, where exhaustive enumeration of the full Boolean function space is still feasible, we compute these quantities for all $2^{2^5}$ functions and computed the empirical joint distribution of $(\text{deg}(f),\textbf{I}[f])$ over the entire function space. We then construct a sample of 2,000 functions by approximately matching this joint distribution: We first grouped functions according to their Fourier degree and Total Influence, and then sampled from each group according to its frequency in the full function space. This procedure ensures that the sampled set of functions spans a broad range of $(\text{deg}, I)$ values, rather than being dominated by structurally random functions, as would occur under uniform sampling of truth tables.

\begin{figure}[h]
    \centering
    \includegraphics[width=0.9\textwidth]{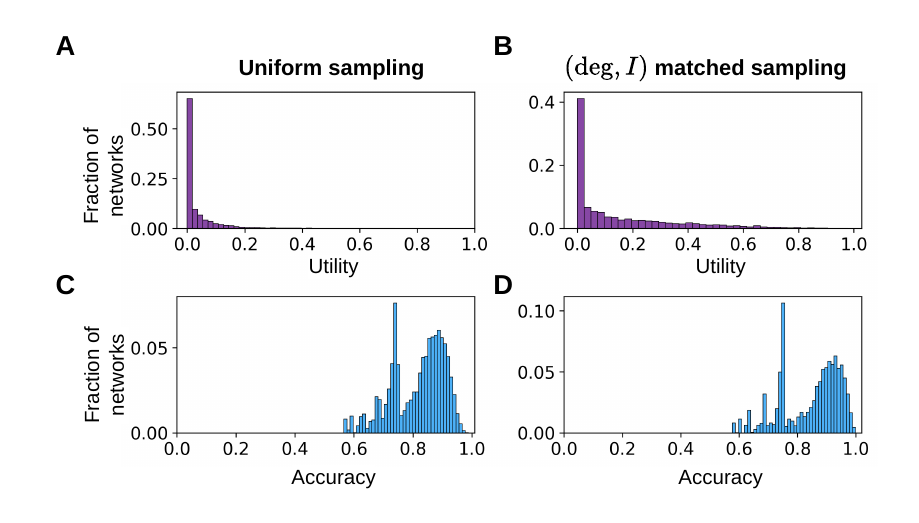}
    \caption{\textbf{Structured sampling of Boolean functions yields network performance distributions very similar to uniform sampling for $N=5$.}
    \textbf{(A,C)} Distributions of Utility (A) and Accuracy (C) across 10,000 Erdős–Rényi networks of size $N=5$ and edge density $0.5$, trained on 2,000 Boolean functions sampled uniformly at random by independent Bernoulli trials with $p=0.5$.
    \textbf{(B,D)} Same as (A,C), but with 2,000 functions sampled by approximately matching the empirical joint distribution of Fourier degree $\text{deg}(f)$ and total influence $\textbf{I}[f]$ over the full $N=5$ function space. }
    \label{fig:sampling_comparison}
\end{figure}

To assess the impact of the sampling scheme, we sampled 10,000 recurrent networks of size $N=5$ from the Erdős–Rényi model (see \nameref{Methods}) with edge density $p=0.5$, and trained each network on both function sets: 2,000 functions sampled uniformly at random (Bernoulli, $p=0.5$), and 2,000 functions sampled by matching the empirical joint $(\text{deg}, I)$ distribution. The resulting Utility and Accuracy distributions are very similar between the two schemes (Fig~\ref{fig:sampling_comparison}). Under both sampling regimes, the vast majority of networks fail to perfectly compute most functions, with Utility concentrated near zero, and Accuracy is broadly distributed and reaches higher values for some networks. The structure-matched sample shifts the Utility distribution slightly: a smaller fraction of networks have near-zero Utility, and the right tail extends further, consistent with the inclusion of more low-degree functions. However, most networks remain incapable of computing most functions exactly. 

We conclude that structured sampling gave little improvement. Moreover, this approach cannot be extended to larger networks as it requires computing $(\text{deg}(f), \textbf{I}[f])$ for all $2^{2^N}$ Boolean functions, which is already infeasible at $N=6$. We therefore used the random sampling for larger networks. 

\newpage 

\printbibliography
\end{document}